\documentclass[a4paper,10pt]{article}
\usepackage[utf8]{inputenc}
\usepackage{graphicx}
\usepackage{amsmath,amsfonts,amssymb}
\usepackage{hyperref}
\usepackage{color}
\usepackage{authblk}

\title{Hard scale uncertainty in collinear factorization: Perspective from $k_t$-factorization}
\author[]{Benjamin Guiot\footnote{e-mail : benjamin.guiot@usm.cl}}
\affil[]{\small{Departamento de F\'isica, Universidad T\'ecnica Federico Santa Mar\'ia; Casilla 110-V, Valparaiso, Chile}}
\date{}

\begin{document}

\maketitle

\begin{abstract}
We analyze two consequences of the relationship between collinear factorization and $k_t$-factorization.
First, we show that the $k_t$-factorization gives a fundamental justification for the choice of the hard scale $Q^2$
done in the collinear factorization. Second, we show that in the collinear factorization there is an
uncertainty on this choice which will not be reduced by higher orders. This uncertainty is absent within the
$k_t$-factorization formalism.
\end{abstract}

\newpage

\tableofcontents

\section{Introduction}
At very high energies, the $k_t$-factorization \cite{ktFac,ktFac2} or semihard approach \cite{semihard1,semihard2} is believed to be the correct formulation.
It is more general than the collinear factorization, and it is well known that the latter is obtained in the $k_t^2\rightarrow0$ limit of the former \cite{ktFac,ktFac2,semihard2}.
If the unintegrated gluon density obeys the BFKL equation \cite{bfkl}, it resums terms proportional to $\alpha_s\ln\left( \frac{1}{x} \right)$,
retaining the full $Q^2$ dependence and not just the leading $\ln Q^2$ terms \cite{ElStWe,revkt}.

One advantage of the $k_t$-factorization is a better treatment of the kinematics, using unintegrated parton densities
and off-shell matrix elements (meaning that the parton virtualities, which can be easily larger than 100 GeV 2 at the
LHC, are not neglected). Moreover, the use of unintegrated parton densities implies that, even at leading order, outgoing partons are not back to back in the laboratory frame.
To obtain this qualitative result with the collinear factorization, it is necessary to go to higher orders (but their computation is easier than in the k t -factorization).
In ref. \cite{revkt}, one can find a table comparing collinear and $k_t$-factorization for different non-inclusive observables.

Here we want to discuss another advantage of the $k_t$-factorization, namely, the disappearance of an uncertainty present in the collinear factorization. In the latter,
the cross section depends on the center-of-mass energy $\sqrt(s)$, the hard scale $Q^2$ , and the factorization scale $\mu$ (and on the renormalization scale, which will be 
ignored in this study). In hadron-hadron collisions, the conventional choice for differential cross sections is $Q^2 \sim p_t^2$ with $p_t$ the transverse
momentum of the outgoing partons in the center-of-mass frame. While current calculations take into account the well-known factorization scale uncertainty, nothing is said
about the choice of the hard scale.

The two main results are given in Secs. \ref{seckt} and \ref{qpt}. In Sec. \ref{seckt}, using the $k_t$-factorization, we demonstrate that
there is an uncertainty in the collinear factorization, coming from the choice of the hard scale. This choice being not
necessary in the case of $k_t$-factorization, the discussed uncertainty is absent in this formalism. Contrary to the
factorization scale, the hard scale uncertainty is not reduced
by higher-order calculations. In Sec. \ref{qpt}, we show that the choice $Q^2 \sim p_t^2 $ can be justified by the dynamical behavior
of the off-shell cross sections and unintegrated parton densities used in the $k_t$-factorization.

\section{Collinear factorization and uncertainties \label{seccol}}

For hadron-hadron collisions, the collinear factorization formula is generally written\footnote{We will use mainly schematic formulas. The sum over parton flavors is ignored 
and one can consider that there is only one flavor (it simplifies also the Dokshitzer-Gribov-Lipatov-Altarelli-Parisi (DGLAP) equation). If not necessary, integrals are not written.}
\begin{equation}
 \frac{d\sigma}{dx_1dx_2dp_t^2}(x_1,x_2,p_t^2,Q^2,\mu^2)=f(x_1,\mu^2)f(x_2,\mu^2)\hat{\sigma}\left(x_1,x_2,p_t^2,\frac{Q^2}{\mu^2}\right).  \label{nlopp}
\end{equation}
The functions $f$ and $\hat{\sigma}$ are the parton densities and partonic cross section, respectively. The variable $p_t$ corresponds to the transverse momentum of outgoing 
partons in the center-of-mass frame. We use the generic notation $Q^2$ for the hard scale which is conventionally identified with $p_t^2$. 
The factorization scale $\mu$ appears due to the renormalization procedure. It comes inside logarithms of the type $\alpha_s\ln(Q^2/\mu^2)$ and has to be chosen close 
to $Q^2$ for an accurate finite order calculation. The dependence on the renormalization scale is not shown, and in this study we take $\alpha_s$ constant. Finally, the longitudinal momentum fractions $x_i$ carried by the incoming partons are given by
\begin{equation}
 x_1=\frac{p_{a,t}}{\sqrt{s}}e^{y_a}+\frac{p_{b,t}}{\sqrt{s}}e^{y_b} \hspace{1cm} x_2=\frac{p_{a,t}}{\sqrt{s}}e^{-y_a}+\frac{p_{b,t}}{\sqrt{s}}e^{-y_b}
\end{equation}
with $a$ and $b$ referring to the two outgoing partons, $y_i$ the rapidities and $s$ the Mandelstam variable for the hadronic system.

The definition of parton densities is not unique \cite{CoSoSt} and for our discussion, it is simpler to shift higher-order corrections from $\hat{\sigma}$ to these these 
functions, leading to the following factorization formula:
\begin{equation}
 \frac{d\sigma}{dx_1dx_2dp_t^2}(x_1,x_2,p_t^2,Q^2,\mu^2)=f(x_1,Q^2;\mu^2)f(x_2,Q^2;\mu^2)\hat{\sigma}\left(x_1,x_2,p_t^2\right).\label{sigQ2}
\end{equation}
Taking into account the first higher-order corrections and following \cite{ElStWe} we write
\begin{equation}
 f(x,Q^2;\mu^2) = f(x,\mu^2) + \frac{\alpha_s}{2\pi}\int_x^1 \frac{d\xi}{\xi}f(\xi,\mu^2)\left(P\left(\frac{x}{\xi}\right)\ln\frac{Q^2}{\mu^2}+C(x)\right), \label{qmu}
\end{equation}
with $C(x)$ a calculable function which is not enhanced by $\ln(Q^2/\mu^2)$. In the following, we will keep the choice and notation of equations (\ref{sigQ2}) and (\ref{qmu}). 

In an all-order calculation, the dependence on the unphysical scale $\mu$ will disappear in both sides of equation \ref{sigQ2}. This is formalized by the DGLAP equation \cite{dgl} (or \cite{ElStWe} for a modern review) which can be written
\begin{equation}
 \frac{ df(x,Q^2;\mu^2)}{d\mu^2}=0. \label{dglap}
\end{equation}
However, in perturbation theory this equation is exact only at a given order, and the parton densities have, in fact, a dependence on $\mu$ 
due to higher-order corrections, justifying our notation. In the opposite, in the r.h.s. of equation (\ref{qmu}), $f(x,\mu^2)$ is given to all orders (by a measurement) 
and depends only on two variables. It is interesting to look at the solution of the DGLAP equation in the $N$-moment space. Inserting (\ref{qmu}) in (\ref{dglap}) 
and taking the Mellin transform we obtain
\begin{equation}
 q^2\frac{\partial f_N(q^2)}{\partial q^2}=\frac{\alpha_s}{2\pi}\gamma_N f_N(q^2) + \mathcal{O}(\alpha_s^2),
\end{equation}
where $\gamma_N$ is the anomalous dimension and, for simplicity, we consider $\alpha_s$ constant. The solution is then given by
\begin{equation}
 f_N(Q^2;\mu^2)=\exp \left[\int^{Q^2}_{\mu^2}\frac{\partial q^2}{q^2}\frac{\alpha_s}{2\pi}\gamma_N \right]f_N(\mu^2)=\left( \frac{Q^2}{\mu^2}\right)^{\frac{\alpha_s}{2\pi}\gamma_N}f_N(\mu^2),\label{FN}
\end{equation}
where we can see the dependence of the parton densities on $Q^2$ and $\mu^2$.\\

\begin{figure}[!h]
\centering
\includegraphics[width=4cm]{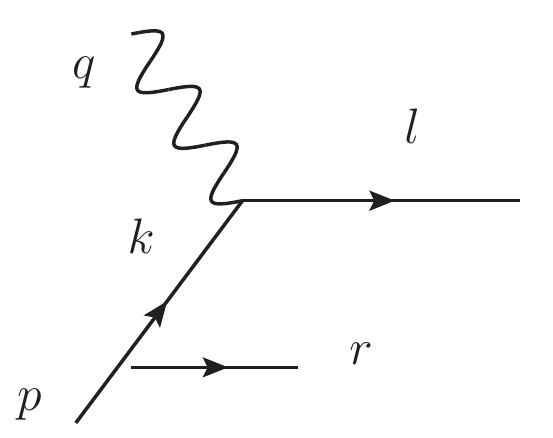}
\caption{Real emission diagram in DIS.}\label{disnlo}
\end{figure}

One remark on the hard scale $Q^2$ before the discussion on uncertainties is in order. The logarithm $\ln(Q^2/\mu^2)$ arises from an integral on the 
transverse momentum, $k_t$, of the incoming parton (an example is given in the case of deep inelastic scattering (DIS) in figure \ref{disnlo}). 
The hard scale appears in the upper bound of such an integral.\\

In this paper, we want to discuss the usual choice\footnote{Or $Q^2=m_t^2$, with $m_t^2=p_t^2+m^2$ and $m$ the mass of the outgoing parton(s).} $Q^2=p_t^2$ done in hadron-hadron collisions, giving the factorization formula
\begin{equation}
 \frac{d\sigma}{dx_1dx_2dp_t^2}(x_1,x_2,p_t^2,\mu^2)=f(x_1,p_t^2;\mu^2)f(x_2,p_t^2;\mu^2)\hat{\sigma}\left(x_1,x_2,p_t^2\right).\label{colfac}
\end{equation}
For definiteness, we consider the case of transverse momentum distribution of heavy quarks in proton-proton collisions at the LHC. This choice for the hard scale means that $p_t^2$ is assumed to be the upper bound for the $k_t^2$ integration. Since, for on-shell partons the kinematical constraint is $k_t^2<\hat{s}/4$ (with $\hat{s}=x_1x_2s$), this is a good approximation in the region $p_t^2\simeq \hat{s}/4$, but it is not correct if $\Lambda_{QCD}^2 \ll p_t^2 \ll \hat{s}/4$. In fact, it is exactly in this region that the
$k_t$-factorization is expected to give important corrections.

In the following, we will argue that the $k_t$-factorization provides a fundamental explanation on why choosing the hard scale to be $p_t^2$ is correct. 
But we will also see that this choice is not unique and gives rise to a theoretical uncertainty (in the collinear factorization case) which is not taken into account in 
current calculations. This uncertainty is not reduced by higher-order corrections. The other uncertainties come from the choice of the factorization scale, the mass and parton densities.

\section{$k_t$-factorization}
The $k_t$-factorization (sometimes called high energy factorization or semihard approach) has been developed in parallel in refs. \cite{ktFac,ktFac2,semihard1,semihard2}. In order to include
all theoretical and phenomenological studies, we define the $k_t$-factorization as a convolution of unintegrated gluon densities with off-shell cross sections. For
hadron-hadron collisions, it can be written
\begin{multline}
 \frac{d\sigma}{dx_1dx_1d^2p_t}(s,x_1,x_2,p_t^2,\mu^2)=\int^{k_{max}^2}d^2k_{1t}d^2k_{2t} F(x_1,k_{1t}^2;\mu^2)F(x_2,k_{2t}^2;\mu^2)\times \\
 \times \hat{\sigma}(x_1x_2s,k_{1t}^2,k_{2t}^2,p_t^2). \label{ktfac}
\end{multline}
The variables $k_{1t}$, $k_{2t}$, $x_1$ and $x_2$ refer to the two
spacelike partons entering in the $2\rightarrow 2$ perturbative QCD process. They correspond to the transverse momentum and the hadron longitudinal momentum fraction. The variable $p_t$ is for the 
transverse momentum of the outgoing parton. The precise definition for the upper bound $k_{max}^2$ will be given in another publication; here it is sufficient to know that
\begin{equation}
 k_{max}^2 > x_1x_2s/4 = p_{t,max}^2.
\end{equation}
In this paper, the only restriction on the unintegrated gluon density $F(x,k_{t}^2;\mu^2)$ is that the second scale $\mu^2$ has to be interpreted as the factorization scale. 
In this case, it is related to the usual gluon density by
\begin{equation}
 f(x,Q^2;\mu^2)=\int^{Q^2}F(x,k_t^2;\mu^2)d^2k_t, \label{undens}
\end{equation}
where we follow the notation used in refs. \cite{CaCiHa,CaCiHa2} \footnote{However our function $F(x,k_t^2;\mu^2)$ is related to their function by a factor $x$.}. For completeness we mention
that for practical purposes, a specific treatment has to be done in the infrared. Some examples can be found in refs. \cite{LiSaZo,ShShSu}. The unintegrated 
gluon density can be obtained by inverting relation (\ref{undens}):
\begin{equation}
 F(x,k_t^2;\mu^2)=\frac{1}{\pi}\frac{\partial f(x,k_t^2;\mu^2)}{\partial k_t^2}.
\end{equation}
The corresponding equation in the $N$-moment space is
\begin{equation}
 F_N(k_t^2;\mu^2)=\frac{\alpha_s}{2\pi^2 k_t^2}\gamma_N \left(\frac{k_t^2}{\mu^2}\right)^{\frac{\alpha_s}{2\pi}\gamma_N}f_N(\mu^2), \label{fn2}
\end{equation}
where equation (\ref{FN}) has been used. Equation (\ref{ktfac}) is for instance valid for KMR\cite{kmr} and BFKL unintegrated gluon densities. In the latter case, an expression for 
$F_N(k_t^2;\mu^2)$ can be found in refs. \cite{CaCiHa,CaCiHa2}. The factor $\frac{\alpha_s}{2\pi}\gamma_N$ is replaced by $\gamma_N(\alpha_s)$ which has a perturbative expansion in $\alpha_s/N$,
first obtained by Balitsky, Fadin, Kuraev and Lipatov\cite{bfkl}.\\

The cross section $\hat{\sigma}$ is computed using off-shell matrix elements (see \cite{revkt} for more details). We will discuss the case where outgoing partons are on-shell and the two incoming partons are spacelike, with off-shellness
$k_1^2\simeq-k_{1t}^2$ and $k_2^2\simeq-k_{2t}^2$. We will see that taking into account this degree of freedom (which requires additional integrations on $\overrightarrow{k_t}$)
is the reason behind the disappearance of the uncertainty on $Q^2$, presented in the previous section.\\

We will close this section by two remarks. The most general expression for $F_N(k_t^2;\mu^2)$ derived in ref. \cite{CaCiHa3} in the $N\rightarrow 0$ limit and minimal 
subtraction scheme is
\begin{equation}
 F_N(k_t^2;\mu^2)=R(\alpha_s) \frac{\gamma_N(\alpha_s)}{\pi} \left(\frac{k_t^2}{\mu^2}\right)^{\gamma_N(\alpha_s)}f_N^{\overline{MS}}(\mu^2),
\end{equation}
with $R(\alpha_s)$ having the following perturbative expansion:
\begin{eqnarray}
 R(\alpha_s)&=&1+\frac{8}{3}\zeta(3)\left(\frac{\overline{\alpha}_s}{N}\right)^3-\frac{3}{4}\zeta(4)\left(\frac{\overline{\alpha}_s}{N}\right)^4+ \nonumber \\
            &+&\frac{22}{5}\zeta(5)\left(\frac{\overline{\alpha}_s}{N}\right)^5+  \mathcal{O}\left(\left(\frac{\overline{\alpha}_s}{N}\right)^6\right)
\end{eqnarray}
with $\overline{\alpha}_s=C_A\alpha_s/\pi$ and $\zeta(n)$ the Riemann zeta function. The expression given for $F_N(k_t^2;\mu^2)$ in equation (\ref{fn2}) corresponds to the 
lowest order ($R=1$).\\

The second remark is that one can encounter the following definition:
\begin{equation}
 f(x,\mu^2)=\int^{\mu^2}d^2k_t F(x,k_t^2,\mu^2).
\end{equation}
By writing the l.h.s. $f(x,\mu^2;\mu^2)$, we can see that this is nothing else than our definition (\ref{undens}) with the choice $Q^2=\mu^2$.

\section{Relationship between collinear and $k_t$-factorization: discussion on the hard scale uncertainty \label{seckt}}
In section \ref{seccol}, we discussed the fact that in the collinear factorization a choice for $Q^2$ has to be done and that it should be accompanied by an uncertainty.
The reason why this uncertainty is absent in the $k_t$-factorization is because the transverse momentum dependence of the incoming partons is explicitly taken into account and 
integrated up to the kinematical upper bound $k_{max}^2$, c.f equation (\ref{ktfac}). It is not necessary to choose the physical scale inside the unintegrated parton densities
since all possibilities are taken into account, ``weighted'' by the $k_t$-dependent off-shell cross section.\\

To understand why, in equation (\ref{colfac}), the scale inside the parton density is approximatively $p_t^2$ and why the  collinear factorization still 
works\footnote{To be precise on this
statement, one should specify the process under consideration. Here we mean that, for sufficiently inclusive quantities, there is no huge discrepancy.} at $p_t^2 \ll s$, 
it is interesting to see how the collinear factorization can be found as a limit of the $k_t$-factorization. Equation (\ref{ktfac}) can be written
\begin{multline}
  \frac{d\sigma}{dx_1dx_2dp_t^2}=\int^{p_t^2}d^2k_{1t}d^2k_{2t} F(x_1,k_{1t}^2;\mu^2)F(x_2,k_{2t}^2;\mu^2)\hat{\sigma}(x_1x_2s,k_{1t}^2,k_{2t}^2,p_t^2) + \\
 \int_{p_t^2}^{k_{max}^2}d^2k_{1t}d^2k_{2t} F(x_1,k_{1t}^2;\mu^2)F(x_2,k_{2t}^2;\mu^2)\hat{\sigma}(x_1x_2s,k_{1t}^2,k_{2t}^2,p_t^2) \label{2int}=I^{cf}+I^{ct}.
\end{multline}
The off-shell cross section is built in order to give the usual on-shell cross section in the limit $k_{it}^2 \ll p_t^2$. Then the first term above can be approximately written
\begin{equation}
I^{cf}=\hat{\sigma}(x_1x_2s,p_t^2) \int^{p_t^2}d^2k_{1t}d^2k_{2t} F(x_1,k_{1t}^2;\mu^2)F(x_2,k_{2t}^2;\mu^2).
\end{equation}
Here $\hat{\sigma}(x_1x_2s,p_t^2)$ refers to the on-shell cross section (since it has no $k_{it}^2$ dependence). Finally, using the definition (\ref{undens}), we obtain
\begin{equation}
I^{cf}=f(x_1,p_t^2;\mu^2)f(x_2,p_t^2;\mu^2) \hat{\sigma}(x_1x_2s,p_t^2).
\end{equation}
Comparing this expression with equation (\ref{sigQ2}), we see that it corresponds to the collinear factorization formula with the choice $Q^2=p_t^2$. 

Splitting the integral at $2p_t^2$ instead of $p_t^2$ will not change anything for the $k_t$-factorization, while the collinear factorization
part will be given by
\begin{equation}
 I^{cf}=f(x_1,2p_t^2;\mu^2)f(x_2,2p_t^2;\mu^2) \hat{\sigma}(x_1x_2s,p_t^2),
\end{equation}
showing that there is an uncertainty on the choice of the hard scale. This procedure, which explains why $p_t^2$ appears in the parton densities, makes sense only if the second term
in equation (\ref{2int}) gives a correction. Formally, there is an uncertainty on the choice of the scale $Q^2$ making the second integral in equation (\ref{2int}) small:
\begin{multline}
I^{ct}=\int_{Q^2}^{k_{t,max}^2}d^2k_{1t}d^2k_{2t} F(x_1,k_{1t}^2;\mu^2)F(x_2,k_{2t}^2;\mu^2)\times \\
 \times \hat{\sigma}(x_1x_2s,k_{1t}^2,k_{2t}^2,p_t^2) \ll f(x_1,Q^2;\mu^2)f(x_2,Q^2;\mu^2) \hat{\sigma}(x_1x_2s,p_t^2). \label{intap}
\end{multline}
Note that $Q^2$ can be interpreted as the effective upper bound for $k_t^2$ integration.\\

To summarize this section, we can rewrite equation (\ref{2int}) as
\begin{equation}
 I()= I^{cf}(Q^2)+I^{ct}(Q^2),\label{conttot}
\end{equation}
with $I^{cf}$ the collinear factorization contribution and $I^{ct}$ a correction term. An appropriate hard scale fulfills the relation
$I^{ct}(Q^2)\ll I^{cf}(Q^2)$, and its choice is not unique. This is the first main result of this paper. The uncertainty on this choice is not taken into account in current 
calculations and could be numerically large compared to the factorization scale uncertainty, the latter being reduced by higher-order calculations. The l.h.s of equation (\ref{conttot}) does not depend on $Q^2$. As explained in the beginning of this section, the hard scale uncertainty is absent in the $k_t$-factorization formalism.

\section{Choosing the hard scale}\label{qpt}

We will now explain qualitatively why $Q^2=p_t^2$ is an acceptable choice, making the collinear factorization formula 
accurate (in the sense that equation (\ref{intap}) is true), even at small transverse momentum. To understand this, we will consider separately the cases of high $p_t$ and low $p_t$ ($\sim 1$ GeV). 

\begin{figure}[!h]
\centering
\includegraphics[width=11cm]{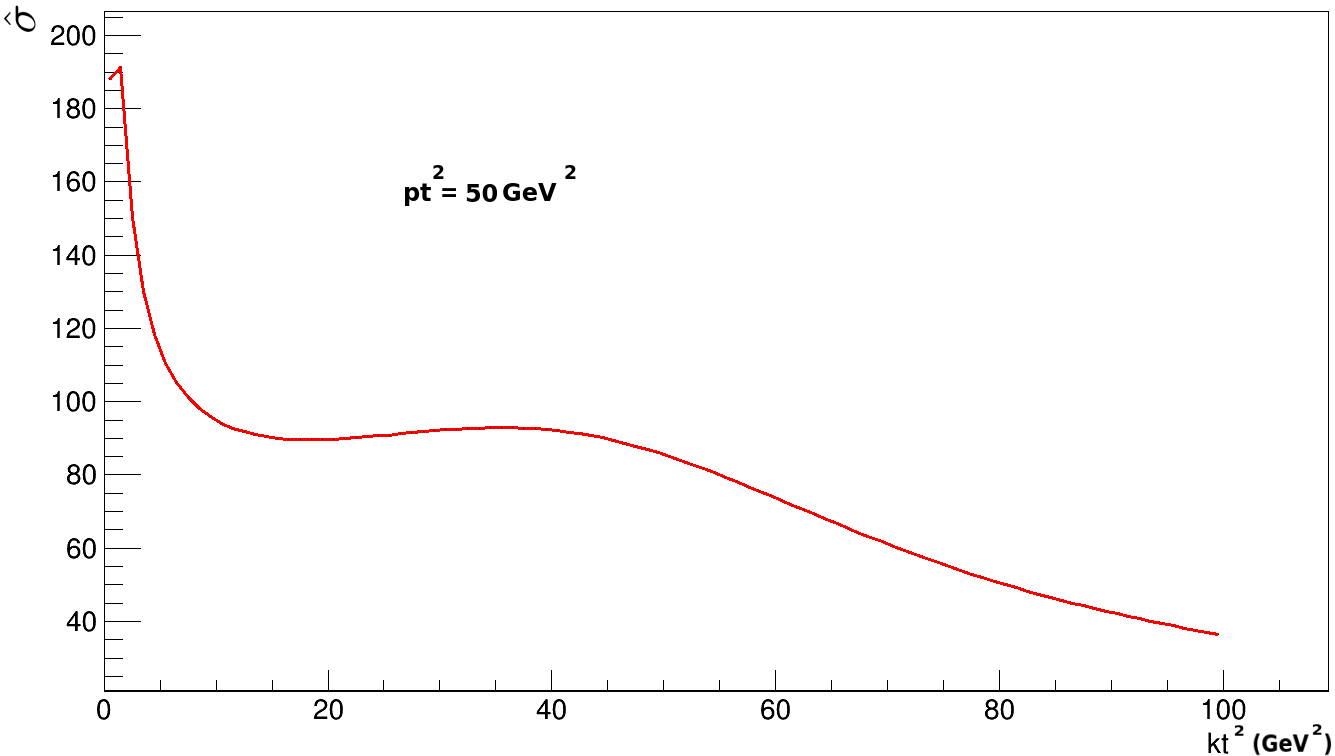}
\includegraphics[width=11cm]{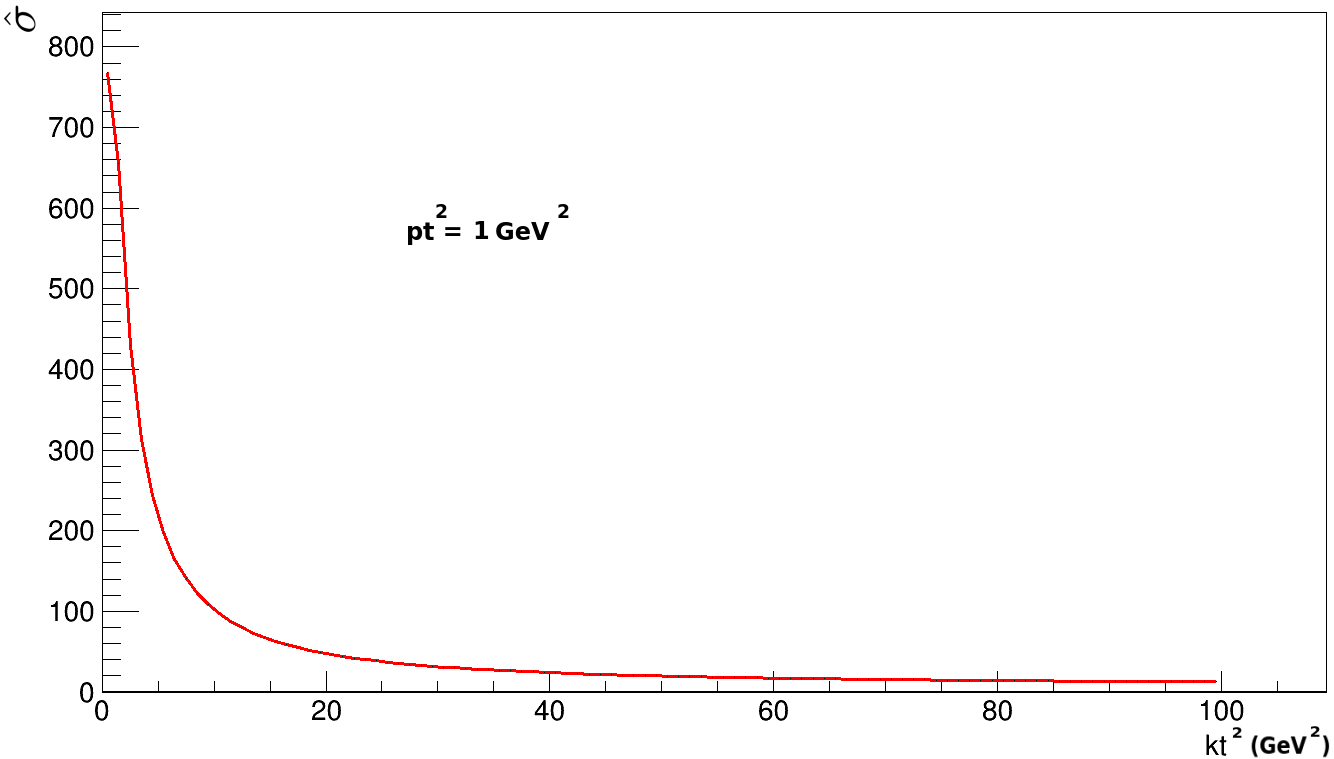}
\caption{Off-shell cross section for the process $gg\rightarrow Q\overline{Q}$ (taken
from ref. \cite{ktFac2}) as a function of the transverse momentum $k_t^2=k_{1t}^2=k_{2t}^2$ of the incoming 
spacelike partons. Top: For central rapidity, $y=0$, and $p_t^2=50$. Bottom: $y=0$, and $p_t^2=1$. Other
variables have been integrated out.}\label{offshcross}
\end{figure}

The reason why the integral $I^{ct}$ can be small even if the phase space for integration is large is due to the fact that in the region $1 \ll p_t^2 < k_{i,t}^2< k_{t,max}^2$ the
off-shell cross section is slowly decreasing with $k_t^2$ (factor 2 between 0 and 40 GeV$^2$; see figure \ref{offshcross}, upper panel), while the unintegrated gluon density is strongly suppressed 
(by power of $k_t^2$)\footnote{At small $x$. For $x>x_0 \sim 0.01$ the suppression is even exponential.}.

Consequently, in the high $p_t$ case, what matters is to integrate up to a large scale, which can be $Q^2=p_t^2$ but also $Q^2=4 p_t^2$. In any case, all this kinematical region is suppressed by
the unintegrated parton densities. The small sensitivity of the result to this scale was expected from  the $\ln (Q^2/\mu^2)$ behavior of partons densities, see equation (\ref{qmu}).\\

At small $p_t$, the suppression due to the unintegrated gluon density is not enough to explain why equation (\ref{intap}) is true if one chooses $Q^2=p_t^2$. 
But, in this region, the off-shell cross section decreases quickly with $k_t^2$ (figure \ref{offshcross}, lower panel), making the integration up to $\sim p_t^2$ sufficient.\\

This is our second main result. The choice $Q^2=p_t^2$ for the hard scale is explained by the dynamical behavior of the unintegrated parton densities and the off-shell cross section. The role of the off-shell cross section
in choosing the effective cut-off for the $Q^2$ integration in DIS has been underlined in \cite{CaCiHa}.

\section{Conclusion}
We have seen that, by choosing correctly $Q^2$ in equations (\ref{intap}), the $k_t$-factorization formula can be split into two parts: the collinear factorization
plus a correction term. In this case, the scale $Q^2$ can be interpreted as the effective upper bound for the $k_t^2$ integration. Based on the behavior of the unintegrated gluon 
density and the off-shell cross section, we argued that in the case of hadron-hadron collisions the choice $Q^2=p_t^2$ can be made, but it is not unique.

Consequently, there is an uncertainty coming from the choice of the hard scale. The difference with the uncertainty on the factorization scale is that it is not reduced by higher-order corrections, the reason being that it does not obey 
a renormalization group equation. This uncertainty is absent in the $k_t$-factorization, thanks to the integration on the transverse momentum.

In arriving at this conclusion, we used quite general arguments, and our results can be easily extended to other cases where the factorization formulas are valid.\\

A practical consequence is that the uncertainty estimation within the collinear factorization is underestimated (usually, the estimation of uncertainties is done for the mass, 
the parton densities and the factorization scale). Note that, instead of $Q^2=p_t^2$, we could choose this scale
in order to keep the correction term at 1$\%$ (for instance). Then we can expect a more complicated relation $Q^2=f(p_t^2)$. In particular, one should have $Q^2>p_t^2$ at small $p_t$,
since this is the kinematical region where the collinear factorization is less accurate.

\section*{Acknowledgement}
We would like to thank F. Hautmann, H. Jung and J. Bartels for useful discussions and valuable comments. We are also grateful to E. Levin for interesting discussions.
We acknowledge support from Chilean FONDECYT grants 3160493.  We acknowledge support by the Basal project FB0821.



\begin{thebibliography}{9}
\bibitem{ktFac} J.C. Collins and R.K. Ellis, Phys. Rev. B360 (1991) 3-30
\bibitem{ktFac2} S. Catani, M. Ciafaloni and F. Hautmann, Phys. Rev. B366(1991) 135-188
\bibitem{semihard1} L. Gribov, E. Levin, M. Ryskin, Phys. Rep. 100 (1983) 1
\bibitem{semihard2} E. M. Levin, M. G. Ryskin, Y. M. Shabelski, A. G. Shuvaev, Sov. J. Nucl.Phys. 53 (1991) 657
\bibitem{bfkl} E. Kuraev, L. Lipatov, V. Fadin, Sov. Phys. JETP 44 (1976) 443\\
               E. Kuraev, L. Lipatov, V. Fadin, Sov. Phys. JETP 45 (1977) 199\\
               I. Balitsky, L. Lipatov, Sov. J. Nucl. Phys. 28 (1978) 822
\bibitem{ElStWe} R. K. ELLIS, W. J. STIRLING AND B. R. WEBBER, “QCD and Collider Physics”, Cambridge university press (1996)
\bibitem{revkt}B Anderson et al. , hep-ph/0204115v2
\bibitem{CoSoSt} J. Collins, D. Soper and G. Sterman, ``Factorization of Hard Processes in QCD'', arXiv:hep-ph/0409313 (2004)
\bibitem{dgl} V.N. Gribov and L.N. Lipatov, Sov. J. Nucl. Phys. 15 (1972) 438-450\\
                G. Altarelli and G. Parisi, Nucl. Phys. B126 (1977) 298-318\\
                Y.L. Dokshitzer, Sov. Phys. JETP 46 (1977) 641-653
\bibitem{CaCiHa} S. Catani, M. Ciafaloni and F. Hautmann, Nuclear Physics B (Proc. Suppl.) 29A (1992) 182-191
\bibitem{CaCiHa2} S. Catani, M. Ciafaloni and F. Hautmann, In *Hamburg 1991, Proceedings, Physics at HERA, vol. 2$^*$ 690-711 and CERN Geneva - TH. 6398 (92/01,rec.Apr.) 21 p
\bibitem{LiSaZo} A.V. Lipatov, V.A. Saleev and N.P. Zotov,  hep-ph/0112114 (2001)
\bibitem{ShShSu} Yu.M. Shabelski, A.G. Shuvaev and I.V. Surnin, Int. J. Mod. Phys. A33 (2017) no.01, 1850003
\bibitem{kmr} M. A. Kimber, A. D. Martin, M. G. Ryskin, Phys. Rev. D63 (2001) 114027
\bibitem{CaCiHa3} S. Catani, M. Ciafaloni and F. Hautmann, Physics Letters B 307 (1993) 147-153
\end{thebibliography}
\end{document}